# TOKEN-BASED INSURANCE SOLUTIONS ON BLOCKCHAIN

| Simon Cousaert | Nikhil Vadgama | Jiahua Xu |
|---|---|---|
| The Block | UCL CBT | UCL CBT |
| scousaert@theblock.co | nikhil.vadgama@ucl.ac.uk | jiahua.xu@ucl.ac.uk |

**Abstract**

With the rising demand for protection against new risks such as loss of digital assets, novel insurance services and products emerge. In particular, token-based insurance solutions on blockchain transform the insurance business by providing cover for new risks and streamlined, (semi-)automated underwriting and claim processes. In the chapter, we present a general framework of token-based insurance solutions, delegating their fundamental building blocks that include core roles, main tokens and assets, as well as key processes and operations. We describe three major token-based insurance solutions in the market and compare them in terms of native token functionality, tokenized cover types, claim assessment process and capital model. Based on the discussion of the general framework and concrete examples of token-based insurance solutions, we summarize their advantages and point out their drawbacks. We conclude that despite being at a nascent stage, the token-based insurance space bears the promise to unseat the incumbent players with more use cases being explored and more user activities.

**Acknowledgements**

The authors thank Niklas Häusle, Vincent Piscaer, Juan Ignacio Ibañez, Matthew Sheridan, and editors Horst Treiblmaier and Mary Lacity for their valuable comments.

**Introduction**

Insurance plays a vital role in dealing with risks and uncertainties in society. It provides us with financial compensation when we suffer from losses caused by various unfortunate events: from health impairment to job loss, from robbery to traffic accidents. As early as two thousand years ago, Indian, Chinese and Babylonian traders practiced methods of pooling risk (Dewan, 2008; Vaughan, 1996). For example, Chinese seafaring merchants pooled together their goods into a collective fund that would pay out if there was any damage to any of the members' ships (CB Insights, 2021). Today, the global insurance industry stands at taking in approximately 6 trillion USD of premiums. To put this number into context, this industry is larger than the overall economy of countries such as Germany and Japan. Despite the long history of insurance in society, the industry is required to constantly evolve to tackle, among other things,

1. the emergence of new risks, especially due to new technology development, such as cyber security breaches;
2. the transformation of existing risks, such as increasingly recurring natural catastrophes due to climate change;
3. shifts in consumer needs, such as rapid access to customized, on-demand insurance, quick claim management, as well as transparency in the insurance processes;
4. continuous advancement in techniques applied in insurance fraud.

Against this backdrop, there is a dire need for solutions that allow insurance policies to be flexibly designable, insurance holder and claim data to be easily manageable, and insurance processes to be openly auditable. Bearing the nature of programmability, traceability and transparency, token-based insurance solutions built with blockchains are on the rise. In particular, the plethora of token models underpinned by smart contracts enables easy configuration of various products and services (Xu & Xu, 2022), as well as cost-efficient record-keeping of miscellaneous transactions and activities within the insurance business.

In the insurance industry, novel projects that utilize blockchain as an infrastructure range from those still at a proof-of-concept phase to those that have already gone into production. As an example of a start-up project, Claimshare (2021) combats double payouts for the same incident at different insurers, an issue that affects 5-10% of insurers payouts. This project has won awards and partnered with Intel and KPMG (CB Insights, 2021). State Farm, the largest auto insurer in the US and the USAA (United Services Automobile Association) are using blockchain to settle subrogation claims between themselves in the property and casualty insurance sector. This system was announced as being in production in January 2021, helping automate and secure a previously manual process, thereby speeding up approximately 75,000 claims exchanged between these two companies (StateFarm, 2021). Finally, industry consortia have also been set up, such as B3i (the Blockchain Insurance Industry Initiative), incorporated in 2018 and owned by 21 global insurance participants—including Allianz, AXA, Generali, Zurich, Swiss Re—with over 40 companies as shareholders, customers, and community members. With the vision for "better insurance enabled by frictionless risk transfer" (B3i, 2021a), B3i is creating insurance products on the blockchain and has released its first application that manages catastrophe excess of loss with a pipeline of new features expected to be added (B3i, 2021b).

The focus of this chapter is the area of novel token-based business models on blockchain. We explore how the insurance business can transform through the use of distributed ledger technology and tokenization. We first lay out the preliminaries of token-based insurance solutions; we then compare in detail three insurance protocols on blockchain, namely Nexus Mutual, Etherisc and inSure; this is followed by a discussion on further challenges faced by token-based insurance solutions; we conclude the chapter with an outlook of the insurance industry in the token economy.

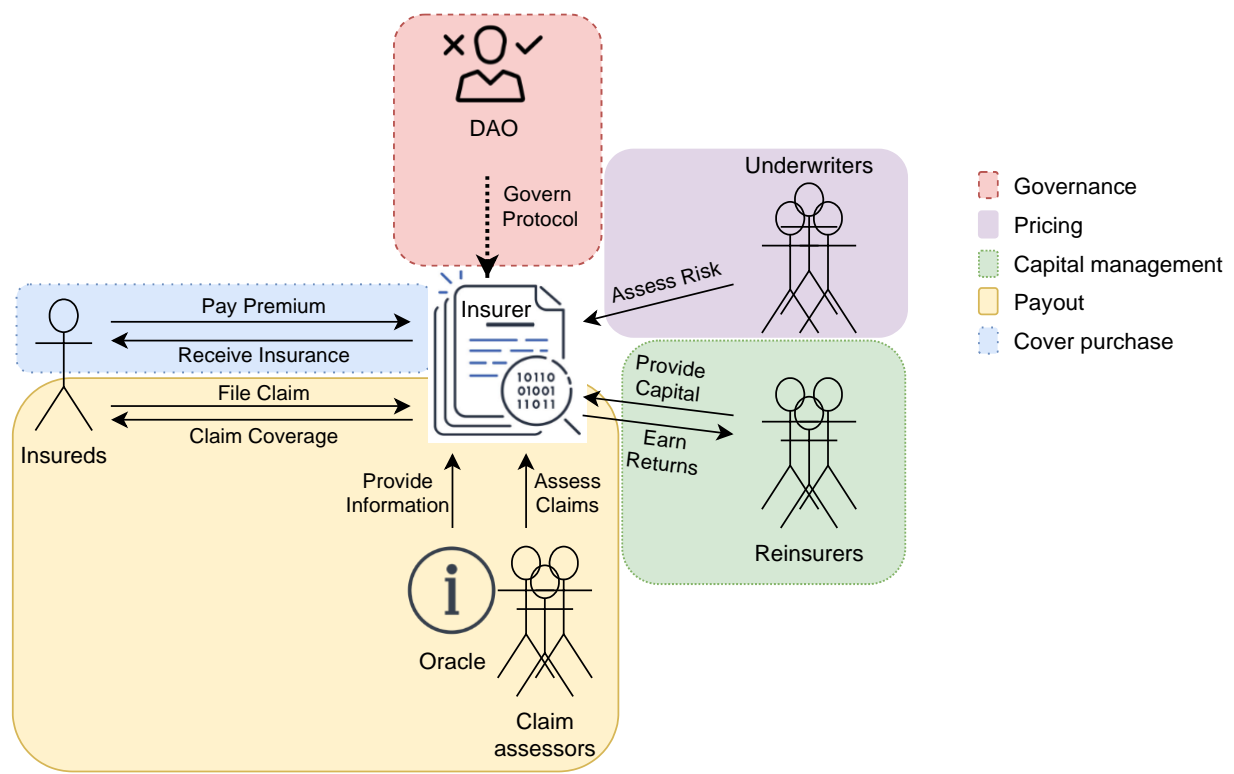

Figure 1: Different actors and actions of a stylized token-based insurance solution

**Preliminaries of Token-based Insurance Solutions**

In this section, we dissect major components of token-based insurance solutions—including core roles, main tokens and assets, key processes and operations—and discuss how those components are associated with each other. Figure 1 shows a stylized illustration of a general token-based insurance solution.

*Roles*

The token-based insurance model requires many actors, each of which has a vital role to play in the ecosystem. The main roles in this interplay include the insureds, the insurers, the underwriters, the reinsurers, the claim assessors and, finally, a decentralized autonomous organization (DAO) responsible for governance.

*Insureds*

Insureds are protection seekers, the target consumers of insurance products and services. They pay premiums—in the form of some cryptocurrency designated by the insurance platform (see section "Different from workflows of traditional insurance, token-based insurance solutions are underpinned by a multitude of digital assets and tokens. A permissionless and public blockchain allows anyone to transact and verify transactions with these tokens. As discussed further in this chapter, the token economics (or "tokenomics") define the utility of each token and the ways the token can be used to incentivize positive behavior in the network.

Payment Token" for further details)—to alleviate the potential negative impact on future hazards. The insureds are financially compensated—again in the form of some cryptocurrency—if the hazard occurs.

*Insurer*

The insurer's role is typically played by the insurance protocol, which consists of smart contracts that can hold premiums paid in tokens and redistribute funds to insureds with approved claims. For a mutual insurance protocol, insureds are simultaneously owners of the protocol and are entitled to the protocols' premium surplus after claim payout deductions.

*Underwriters*

Underwriters assess risk levels of uncertain events. If empirical data on the occurrences of a risk event is available, conventional actuarial models are applicable for risk evaluation. This computation can be performed automatically by the protocol itself. In this case, the same party can undertake the insuring and underwriting roles: the protocol. Absent historical statistics, underwriting can be achieved through the wisdom of the crowd. In this case, a group of platform users can express their view on the riskiness of an event by staking their tokens with a quantity of their choice. Should the risk event occur, their stake will be contributed to the cover; otherwise, the users will be awarded, proportionate to their stake, with part of the protocol's premium income.

*Reinsurers*

Absent reinsurers, an insurance protocol's payout capacity is limited to premiums received. Reinsurance increases an insurance protocol's solvency through an additional layer of financial guarantee. In the decentralized space, the role of the reinsurer is typically assumed by risk-taking investors, who likely share part of the protocol's profit as a form of reinsurance premium, but, expectedly in rare cases, can also suffer losses (Karpischek et al., 2016). In that sense, users who stake tokens on risk events can also be deemed to be reinsurers to a certain extent. Indeed, they inject additional capital to the insurance payout pool, and suffer loss when a certain level of risk events is exceeded but earn returns otherwise. With the contribution of their funds, they reduce the capital required by the other contributors, thereby reducing the insurer's risk exposure.

*Claim Assessors*

For certain insured risks, a smart contract can automatically assess the legitimacy of claims by reading data feeds provided by oracles. It is possible to distinguish between centralized oracles and decentralized oracles. Centralized oracles are controlled by a limited number of entities. Data can be fed in through manually emitted transactions or automatically through cross-platform protocols such as Chainlink (2021). Decentralized oracles are typically operated through on-chain protocols, such as price oracles provided by decentralized exchanges with automated market-making protocols (Xu et al., 2022). For claims that are difficult to evaluate automatically, the assessors can be a group of users who express their approval or objection towards the claim with the voting power proportionate to their token holding (usually designated governance tokens). With the voting method, a well-designed incentive and penalty scheme must be in place to encourage honest assessment and deter malicious vote manipulation (Braun, Häusle, et al., 2022).

*Decentralized Autonomous Organization*

Many new token-based insurance solutions have a decentralized autonomous organization (DAO) as a key component in governing the protocols. DAO is a collective noun referring to everyone eligible to participate in a protocol's governance. Eligibility is often represented by the holding of governance tokens. The decentralized nature of tokens means that no single entity controls the mechanics and economics behind a token. Still, those decisions are made by a community organized around the set of rules imposed by the token (Cointelegraph, 2021). The DAO operates using smart contracts that establish the rules. People with a stake in the DAO are often rewarded with voting rights, frequently in the form of a governance token. Thus, a governance token holder often has a say in how a DAO-governed protocol should change elements like protocol parameters, budgeting, and treasury expenditures. In its ultimate form, a change to the DAO is proposed by a community member and voted upon by the governance token holders, after which the proposal gets implemented or not. A DAO can consist of the insurers, insureds, underwriters, claim assessors, and governance token holders.

*Assets and Tokens*

Different from workflows of traditional insurance, token-based insurance solutions are underpinned by a multitude of digital assets and tokens. A permissionless and public blockchain allows anyone to transact and verify transactions with these tokens. As discussed further in this chapter, the token economics (or "tokenomics") define the utility of each token and the ways the token can be used to incentivize positive behavior in the network.

*Payment Token*

Most protocols have a designated token for premium payments. The designated premium token can be the same as the denominating currency of a potential payout (e.g. SURE in InsurToken). If they differ, the insureds can often purchase cover in a desired payout currency, usually a widely-adopted cryptocurrency such as ETH or DAI (e.g. Nexus Mutual). For user-friendliness, some protocols (e.g. Etherisc) also allow cover to be purchased directly in fiat currency. While different currencies can be accepted, some protocols (e.g. Nexus Mutual) convert all payments in the back-end into a designated token for the ease of accounting, mostly via a decentralized exchange (BraveNewDeFi, 2021b).

*Insurance Token*

An insurance token represents a certificate of cover obtained by the insured in return for the premium payment. The insurable risk covered by a policy is usually non-transferable, and hence naturally, undividable. Therefore, an insurance policy can be represented by a non-fungible token (NFT), corresponding to the exact risk covered by the policy. However, not every insurance protocol employs insurance tokens. Some protocols rely on immutable on-chain transaction records as proof of cover (such as Nexus Mutual), while others implement hybrid bookkeeping systems with a central database to record insurance purchases. Despite the indivisibility of insurable risk, the beneficiaryship of a policy may be fractionalized and transferred. As such, a tokenized insurance policy would be easily tradable on a blockchain-based secondary market (e.g. fidentiaX, 2021).

*Governance Token*

In line with decentralization principles, blockchain-based insurance protocols typically employ governance tokens for a distributed sovereignty. First and foremost, governance token holders have voting rights on issues related to protocol-level rules, such as updating the pricing scheme or adding a new insurable risk. Certain protocols also grant governance token holders the voting rights to approve or reject a claim. Governance tokens sometimes also represent ownership of the protocol, where token holders share the profit and loss of the protocol. This representation can be achieved by algorithmically setting the price of the tokens such that the market capitalization of the governance token corresponds with the total funds locked in with the insurance pool contracts (e.g. Nexus Mutual). Governance tokens are sometimes also used for payments, which automatically makes insureds governance token holders, thereby entitling them to a share of the protocol ownership that changes depending on premiums paid and insurance payouts received. Mutual insurance protocols often take this approach.

*Processes and Operations*

The workings of a specific solution are tied to its processes and operations. The unique selling proposition of a protocol is often embedded in the different use cases that it offers to customers. This section seeks to generalize a set of processes and operations frequently observed in token-based insurance solutions.

*Cover Purchase*

Insureds can purchase cover against risks using tokens or currencies permitted by the protocol for premium payment (see section "Different from workflows of traditional insurance, token-based insurance solutions are underpinned by a multitude of digital assets and tokens. A permissionless and public blockchain allows anyone to transact and verify transactions with these tokens. As discussed further in this chapter, the token economics (or

"tokenomics") define the utility of each token and the ways the token can be used to incentivize positive behavior in the network.

Payment Token"). Insureds can customize the cover by specifying parameters such as the amount insured and coverage period. Depending on the protocol specifications, insurable risks may include:

*Financial risks* Insurance can cover negative price movements of a crypto asset below a certain reference price. When purchasing such a type of insurance along with the crypto asset, a user is buying the equivalent of a put option where the underlying asset's price is floored.

*Security risks* Insurance can cover malicious attacks targeting protocols that are underpinned by smart contracts. This type of coverage can provide insurance against the situation where a hacker discovers and exploits a security loophole of a decentralized finance (DeFi) protocol, moving funds from the protocol's smart contract to their address and causing losses on the side of protocol users.

*Off-chain hazards* Insurance can cover unanticipated, unfortunate events such as flight delays. Oracles are needed to feed data of events that are external to the blockchain into the smart contract (Braun, Häusle, et al., 2021).

### *Pricing*

Depending on the protocol, premiums are set either actuarially using historical data or based on the community's collective view. In the case where users stake tokens to signal their belief on the risk level (see section "Underwriters"), the less risky a user believes a contingency to be, the more tokens the user is willing to stake. A higher number of tokens staked in a risk pool reduces the cover price, which correspondingly reflects a lower risk level. In short, pricing of an insurance cover pivots on underwriters' work: all other things equal, the riskier the underwriting result of an event shows, the higher the premium. In addition, the price of a particular insurance cover is also positively correlated with the total amount insured and the coverage period.

### *Payout*

Insurance payouts can either be triggered automatically or processed manually through individually filed claims. In the former case, once information on the occurrence of the events is fed into the insurance smart contract, the payout function is activated, and the affected insureds are immediately compensated. For off-chain risk events, the provision of external data is done by oracles. In the latter case, the legitimacy of a claim is determined with human judgement by the protocol team or a group of claim assessors and through a poll with all eligible protocol participants.

*Capital Management*

Insurance protocols can have a capital model that determines the minimum capital to be held, a metric often set to similar standards as EIOPA's Solvency II (EIOPA, 2020), which ensures a confidence level of 99.5% in solvency over one year. Generally, there are two types of cash flows. First, insureds purchase coverage by paying a premium to a capital pool and the capital pool pays customers in case there is an accepted claim. Secondly, the funds in the capital pool can be invested to generate returns (BraveNewDeFi, 2021a). In the unlikely event that the reserves in the capital pool are completely depleted, capital from other sources must be employed. The governance token holders would likely make this decision. Out of the six biggest insurance protocols listed on Coinmarketcap (2021), there have not yet been any known instances of insolvency and no concrete procedure of how to handle such a situation has been documented.

*Governance*

A governance token holder has a say in how to shape the protocol. In its ultimate form, a change to the protocol is proposed by a community member and voted upon by governance token holders. A decision is never made by only a select group of developers but rather by the whole community that holds voting rights. Governance token holders can take part in underwriting, assessing and deciding on larger changes to the entire insurance protocol.

**Comparison of Existing Decentralized Insurance Protocols**

This section compares three major insurance protocols by market capitalization: Nexus Mutual, Etherisc and inSure (Table 1).

| Protocol | Native token | Market cap (million USD) | Native token functionality | Tokenized cover | Claim assessment | Capital model |
| --- | --- | --- | --- | --- | --- | --- |
| Nexus Mutual | NXM | 815.8 | - Governance<br>- Cover purchase<br>- Claim assessment | No | Decentralized, through NXM holders | Based on EIOPA Solvency II, 99.5% confidence |
| inSure | SURE | 108.4 | - Governance<br>- Cover purchase<br>- Claim assessment | Yes, SURE token itself | Centralized and decentralized, through auditing companies and SURE holders | Based on EIOPA Solvency II, 99.9% confidence |
| Etherisc | DIP | 25.0 | - Reward risk taking<br>- Monetizing services | Yes, via Risk Pool tokens | Depends on product | Based on EIOPA Solvency II, 99.99% confidence |

Table 1: Comparison between major token-based insurance solutions. Market cap equals native token price times its circulating supply, data retrieved on 31 August 2021 from https://coinmarketcap.com/.

The dominant market position of these protocols indicates a certain level of industry recognition. With a heterogeneous pool of actions, these protocols each have their unique take on a generalized framework.

*Nexus Mutual*

Nexus Mutual is a decentralized insurance protocol. The platform token, NXM, has several functionalities, including governance, the ability to purchase cover and the ability to vote on claims. The first version of the application went live on the Ethereum mainnet in May 2019. As of August 2021, the protocol has an active cover amount of over 500 million USD (Nexus Mutual Tracker, 2021). Customers of Nexus Mutual can insure themselves against three main types of risks: the risk of a yield token losing its peg ("yield token cover"), the risk of protocol failures ("protocol cover") and the risk of hacks or halted withdrawals on exchanges or custodial wallets ("custody cover"). Customers can pay in ETH or DAI to buy a cover, which gets automatically converted to NXM by the smart contracts. Bought covers are not tokenized but kept by a central system. If a customer suffers a loss and believes this loss qualifies for the bought cover, they can file a claim. The cost of filing a claim is integrated into the initial amount paid for the cover. After the claim is correctly filed, the outcome is determined by "claim assessors", NXM holders who stake a portion of their NXM. Assessors acting in good faith are rewarded. Those who act fraudulently can be punished by having part of their staked NXM burned. With the help of "Incentivai", a machine learning algorithm that checks user behavior, the advisory board of Nexus Mutual determines whether or not an act is "fraudulent" (Nexus Mutual, 2021).

A final functionality of the NXM token is its use in a process called "risk assessment". "Risk Assessors" can stake NXM in a specific cover of which they believe the risk of an eventual payout is small. This extra capital decreases the premium for that specific cover. Through staking, Risk Assessors earn proportional rewards in NXM equivalent to 50% of the cover premium paid by insureds. This way of incentivizing investors allows the Nexus Mutual protocol to attract more capital. In the event of a payout, the Risk Assessor loses some or all of their deposit. The protocol capital requirements are determined by a capital model that borrows heavily from EIOPA's Solvency II (Karp & Melbardis, 2017). The price of NXM is determined depending on the reserves in the capital pool and the required funds, which in turn affects the whole ecosystem.

*Etherisc*

Etherisc is a protocol that allows anyone to create their insurance products by providing common infrastructure, product templates and insurance license-as-a-service. At the moment, two insurance products have been launched: Crop Insurance and Flight Delay Insurance. Other applications such as Hurricane Protection and Crypto Wallet Insurance are either in the design or prototype phase (Etherisc, 2021). The platform token, DIP, is a core element in the overarching protocol and incentivizes and rewards platform users for bringing risk to the network and building and maintaining products. Specifically, the token is used to buy insurance products, reward users for updating risk models (similar to the Risk Assessors in Nexus Mutual), reward those that provide reliable data/oracles, and reward a multitude of other stakeholders. Next to the platform token, Etherisc introduces Risk Pool Tokens, the tokenized versions of insurance products, allowing a token marketplace where participants can purchase and redeem tokenized covers. The combination of a "risk pool" and a "reinsurance pool" ensure that long-tail events are covered and that there is always enough capital to pay claims. The required capital is also calculated based on the EIOPA Solvency II model, but Etherisc imposes a 99.99% confidence level instead of 99.5%.

*inSure*

inSure is a decentralized insurance protocol that focuses on protecting traders' portfolios against scams, devaluations and stolen funds. The company launched its first features in 2019 and is currently fully operational and progressing on making the assessment and payment processes fully automated. The platform token, SURE, has multiple functionalities, including governance, claim assessments and the ability to purchase cover. Customers of inSure buy cover by buying SURE tokens. This model differs from the Nexus Mutual cover, where one buys specific covers with the NXM token. SURE tokens are the cover, which means they are transferable. The insured buys a certain number of SURE tokens to insure oneself corresponding with a specific inSurance plan (inSure, 2021a). After acquiring and holding SURE tokens for seven days in their wallet, the insured is covered against multiple risks up to a certain level, depending on the insurance plan. If a customer's portfolio is affected by a scam, devaluation or stolen funds, they can file a claim. That claim is voted upon by auditing companies and SURE holders if there is a challenge to the consensus of the auditing companies. It is not clear whether these parties are incentivized for correct behavior. As of August 2021, there remains a lack of publicly available documentation on inSure's capital model. There is a capital pool and a "surplus pool". The capital pool's goal is to support business development. The surplus pool accrues money by collecting premiums paid by customers and is intended to cover all claims. When the surplus pool cannot cover all claims, the capital pool will be used to cover the shortfall.

**Other Insurance-related Solutions on Blockchain**

Besides the peer-to-peer coverage applications listed in the previous section, other protocols also have a relation to the insurance industry, but are not deemed to be insurance specific in the conventional sense. Example applications are insurance marketplaces, financial markets that allow hedging positions and protocols offering protection to borrowers against liquidation of collateral.

*fidentiaX*

fidentiaX is a marketplace for life insurance trading on blockchain. The company raised funds through an ICO (initial coin offering) in late 2017, where the platform's native FDX were distributed to investors in exchange for ETH. On fidentiaX, insureds list their life insurance policy that interested buyers can bid for. The winning buyer pays the insured a lump sum while the insured transfers the beneficiaryship of the policy. After the transaction, the buyer takes over the premium payment and is entitled to receive the maturity benefit of the insurance policy (Braun et al., 2020). The platform is still in the prototype phase. As per the design, three types of tokens are intended to be employed on the platform: (1) ERC-20-based tokens to represent insurance policies (2) FDX for bidding and (3) stablecoin for payment.

*Opyn*

Options in financial markets can be utilized for multiple purposes, but in the context of insurance they function as instruments that can hedge risk. Opyn is a DeFi option and insurance platform that provides protection and hedging instruments for DeFi deposits and ETH risk. The first version of the Opyn protocol offered protection for USDC and DAI deposits on Compound (Perez et al., 2021). The current second version of the Opyn protocol allows anyone to buy, sell and create options on any ERC20 asset. These options are cash-settled European options, whose exercise date coincide with their expiration date. Opyn option products can be created with oTokens, an ERC20 compatible contract that represents the option product. The protocol states that over 1 billion USD notional volume of options have been traded with over 20,000 trades. Its core developers and investors currently govern the platform.

*Mero*

Mero (formerly Backd) provides assurance to users of lending protocols such as Compound and Aave (Xu & Vadgama, 2022). Mero describes itself as "reactive liquidity", preventing undercollateralized loans from becoming liquidated. Mero aims to increase the capital efficiency of asset borrowing in DeFi, where borrowers do not need to exceedingly overcollateralize their loan position in fear of automatically triggered liquidation. Instead, borrowers delegate fund management to the Mero protocol, which tops up collateral whenever a loan position is on the verge of being liquidated, while using the excess capital for yield farming, an investment activity in DeFi (Cousaert et al. 2022).

**Discussion**

Still, in its infancy, the token-based insurance space is very early within its evolution. While the blockchain industry is witnessing a huge wave of interest due to the booming DeFi industry, DeFi projects are currently dominated by lending protocols and decentralized exchanges. Nexus Mutual, with a total value locked (TVL)—the quantity of tokens secured in the DeFi application's smart contracts—of half a billion USD as of August 2021 (DefiLlama, 2021), prides itself as the largest token-based insurance project but ranks only 24th of all DeFi projects by TVL. Despite the immaturity of the token-based insurance sector and the lack of attention received relative to other DeFi areas, many benefits can already be seen. Indeed, some challenges remain unsolved and have not been tested. In this section, we briefly explore these.

*Benefits*

*Efficient Risk Transfer*
Conventionally, systemic risks can be transferred to the capital market: for instance, exposure to catastrophe risk can be packaged and sold in the form of CAT bonds. Similarly, risks can be easily passed on to the capital market with tokenization and thus reduce or even render moot capital requirements on the insurer's side. By shifting the risk completely onto the capital markets, blockchain-based insurance protocols can act solely as a risk bundling and dividing machine, thus evading regulation applied by conventional insurance companies. Tokenization also makes secondary sales easier. If a purchased cover is tokenized, customers can create a secondary market for insurances. In that case, buying and selling commonly issued covers can be handled outside of a specific ecosystem. A centralized layer on top of issuing the covers complicates the selling and transferring of insurance for those who want to monetize their policy or simply move their funds or activities to another address. Tokenizing purchased cover would address these issues. Fungible tokens can be used to set up a secondary market on a decentralized exchange such as Uniswap for people to trade their purchased cover. If the cover is highly personalized or unique, a non-fungible token will offer similar advantages, although a more specialized market will be needed.

*Transparency*

In case all aspects from the purchase of cover to claims and payout are on a blockchain means the entire insurance process is visible end-to-end. This transparency enables participants in the ecosystem to investigate for themselves whether a protocol is trustworthy. Indeed, the growth and success of a protocol depend on good behavior from its community concerning the entire underwriting and claims assessment processes—in stark contrast to traditional insurance products where this trajectory is opaque, which often leads to poor customer experience. Oracles are also utilized to provide relevant information for the assessment of claims. These oracles can also be decentralized in nature, meaning that there is further transparency in how information relevant to a claim is handled, where this information has come from and what relevant references have been used.

*Security*

Security within a token-based insurance protocol is strengthened through several technological and systematic features. Starting with technological features, without discussing the security features of a blockchain system and smart contracts in detail—one of the main points is that the working code of these protocols is visible for anyone to audit. This transparency can help create a secure system whereby faults in a protocol are easily spotted and corrected. Such an environment also inspires trust in a protocol itself as users (who can read the code) are confident in the performance and behavior of the system. With respect to security in the sense of trusting the process of underwriting and claims management—the Decentralized Autonomous Organizations (DAO) that manage these protocols are composed of governance token holders who are able to vote on the management and direction of the protocol. This decentralized approach can help improve the ecosystem's security as the community is incentivized to act in the best interests of the protocol to support an increasing token price. A secure protocol features a good design with a strong incentive for correct reporting and a strong disincentive for fraudulent behavior. As such, whenever an incident occurs, and a customer files a claim, holders of the platform token can use their vote to decide whether that claim is justified and if that incident falls under the bought cover.

The security of a protocol is dependent on the token design that creates functionalities to optimize the operational efficiency of the service. For example, the token can be used to incentivize correct behavior and disincentivize fraud, reward participants for the correct pricing of the insurance services, or reward people outside of the network for providing reliable data through oracles or manual processes.

*Customization & Agility*

Tokens can have many functions programmed into them, which do not rely on single entity intermediaries for overseeing and maintaining those functions, as they are automatically enforced by the smart contracts which can be updated or enhanced through community governance processes. In the context of insurance protocols, tokens and smart contracts can be used to improve cover customization, connecting insurers and insureds faster. In addition to customization, there is a great deal of agility in how insurance protocols can deal with claims through their community governance systems, and indeed agility in the direction of the protocol itself. Insurance protocols operating as DAOs with governance tokens can upgrade themselves, thus altering the practices for underwriting and claims through invoking voting of governance token holders to pass decisions. This gives these new protocols agility in evolving and responding to changes in their markets in a way that traditional insurance players just cannot do.

*Challenges*

**Lack of contingency plan for insolvency**

Existing protocols do not seem to have any contingency plan for insolvency in place. Although most protocols follow the Minimum Capital Requirements (MCR) standards from EIOPA's Solvency II to a certain extent, there is simply not enough data for confident actuarial modeling of the risks insured. According to the UK Financial Services Compensation Scheme (FSCS), there have been 38 insurance defaults within the traditional insurance sector since 1985 that the FSCS has been involved with (FSCS, 2021). In the UK, if an insurer becomes insolvent, the FSCS will cover up to 90% of an insured's claim (subject to certain limits and eligibility of the policy). In addition to protecting insureds through government insurance schemes, an efficient reinsurance market enables traditional insurance players to diversify risks. Within the token-based insurance sector this reinsurance or government backed insurance does not exist and contingencies from protocols have not been elucidated.

*Low Competition and Participation*

One could argue that there is a monopoly player within this sector: Nexus Mutual stands as the largest player, with other projects not having seen many projects built on top of their protocols (e.g., Etherisc with only Crop Insurance and Flight Delay Insurance) or with little participation (e.g. inSure). Between March to August 2021, only ten insurance claims were raised on inSure (inSure, 2021b). Out of these, only one was approved. The other nine were rejected due to not meeting policy requirements. Furthermore, the participation in voting was extremely low, with many claims not receiving any votes. These protocols are indeed new and evolving. In the case of Etherisc, many new insurance products have been designed and are awaiting licensing. In the case of inSure—their roadmap still includes several features to be built to increase participation in the protocol (running through to 2022).

*Fragile Incentive and Penalty Mechanism*

Just like insurance companies, a DAO exhibits a reluctance to approve claim requests. Understandably, a DAO has the incentive, at least in the short run, to retain funds within the protocol whenever possible instead of disbursing them to cover an individual's loss. In the long run, one may argue that persistent reluctance to approve claims will result in diminishing attraction of the protocol to end users, which will subsequently lead to a depreciation of protocol tokens and thus a devaluation of a DAO's token holding. This negative consequence would discourage a DAO from blindly rejecting all incoming claims to ensure the user base, and hence the value of the protocol. Therefore, theoretically, at equilibrium, a DAO should exert sufficient discretion to scrutinize claims. On the one hand, opportunistic and fraudulent claims are screened out to ensure the profitability of the protocol. On the other hand, legitimate claims are needed to get compensated to ensure users' welfare. Unfortunately, current insurance protocol users have exhibited a short-term orientation which the pseudonymous nature of blockchain applications may have exacerbated. Fraudulent behavior is also an issue from both claimants and voting members. Looking at inSure, several claims filed incorrect reference events that did not occur. On Nexus Mutual, there is at least one incident where an insured submitted a false claim and then subsequently voted for it (Nexus Mutual, 2021).

*Centralization*

By definition, a decentralized protocol should be designed to operate, evolve, and grow even after the original development team and the protocol foundation has disbanded. However, current insurance protocols exhibit centralization in many ways. Firstly, most protocol teams take full charge in either the pre-screening of claims or the ultimate judgement on the legitimacy of claims, representing a concentration of power and a single point of failure. Secondly, existing voting schemes might trigger the so-called Matthew effect. Specifically, users who hold a lot of wealth will influence the voting result to their favor, compounding the accumulation of wealth and power. Protocols as such are thus vulnerable to majority attacks, where a user or a group of colluding users holding a significant proportion of the governance token supply can submit false claims and vote in favor of them, pocketing profit from the protocol.

**Conclusion and Outlook**

In this chapter, we have described how many players are using blockchain technology and tokens to create novel insurance products that utilize new technology and business models to cover new types of risk and improve customer experience. Tokens seek to facilitate flexibly designed and openly auditable insurance policies, which is in stark contrast with traditional insurance products. However, the new token-based insurance industry is at a nascent stage. We have discussed many of the benefits and challenges that face this fledgling industry, and the success or failure of this industry ultimately depends on secure and sound protocol design and practical tokenomics.

In the current stage of the industry, there have been many protocols and DeFi projects that have come to the fold but have quickly disappeared, either due to poor and insecure protocol design or simply due to business models where the tokenomics have not garnered a community to develop them. Nexus Mutual is the leading project in the space, and others are still slowly emerging. As interest in the whole DeFi sector increases and as the exchange and lending sub-sectors mature, the insurance sector is likely to gain increased attention. Nexus Mutual boasted over 1.2 billion USD in cover at its peak in 2021 and currently has active cover of approximately 200 million USD out until January 2022. For a new platform with a niche insurance product, these are promising numbers. The Minimum Capital Requirements have also been greater than 100%, reaching as much as 200% at times in 2020. Of the 95 claims made to date, 18 of these were accepted and paid out over 10 million USD (Nexus Mutual Tracker, 2021). Although these claims currently relate to DeFi and Smart Contract risk, more conventional consumer use case products are also developing and slowly emerging, as we have seen in the case of Etherisc. As the DeFi movement increases in size, more attention from new adopters will turn to these new insurance products as well.

Finally, the blockchain itself and the use of tokens have seen use in the traditional insurance sectors, with many proof-of-concept projects taking place and being delivered to consumers. This development is still at an early stage, and inevitably, as the exploration and use of blockchain as an infrastructure in the traditional insurance industry increases, so will the use of tokens. This moment may be the inflection point for the token-based insurance industry to challenge traditional insurance markets.